\documentclass[journal=nalefd,manuscript=letter,usetitle=true]{achemso}
\usepackage{amsmath} 
\usepackage{textcomp} 

\author{Christopher E. Kuklewicz}
\author{Ralph N. E. Malein}
\affiliation{Institute of Photonics and Quantum Sciences, SUPA, Heriot-Watt University, Edinburgh EH14 4AS UK}
\author{Pierre M. Petroff}
\affiliation{Materials Department, University of California, Santa Barbara, CA 93117 USA}
\author{Brian D. Gerardot} 
\email{b.d.gerardot@hw.ac.uk}
\phone{+44-131-451-8069}
\affiliation{Institute of Photonics and Quantum Sciences, SUPA, Heriot-Watt University, Edinburgh EH14 4AS UK}

\title
{Electro-elastic tuning of single particles in individual self-assembled quantum dots}


\begin{document}

\begin{abstract}%
We investigate the effect of uniaxial stress on InGaAs quantum dots in a charge tunable device. Using Coulomb blockade and photoluminescence, we observe that significant tuning of single particle energies ($\approx$ -0.22~meV/MPa) leads to variable tuning of exciton energies (+18 to -0.9~{\textmu}eV/MPa) under tensile stress. Modest tuning of the permanent dipole, Coulomb interaction and fine-structure splitting energies is also measured. We exploit the variable exciton response to tune multiple quantum dots on the same chip into resonance.
\end{abstract}

Precise engineering of particle wave functions for desired applications is a universal challenge in nanotechnology. One common motivation for nanostructure engineering is to create resonance between an optical transition in a quantum emitter and another quantum system, e.g. a cavity mode for cavity QED experiments \cite{Benson2011} or another transition in a separate independent atom-like system for projective measurement based protocols in quantum computing \cite{kok2010}. Another example application, specific to self-assembled quantum dots (QDs), which places stringent conditions on single particle wave functions is the generation of entangled photon pairs via the biexciton $\rightarrow$ exciton $\rightarrow$ ground state cascade \cite{Akopian2006, Bennett2010, Mohan2010, Dousse2010}. Asymmetry \cite{Singh2010,Gong2011} in a typical self-assembled dot leads to a splitting in the intermediate X$^{0}$ state on the order of 10~{\textmu}eV, which is referred to as a fine-structure splitting (FSS). The FSS generally destroys the prospect for entangled photon pair generation.

For a QD, the intrinsic single particle energies and wave functions are dictated by its size, shape, and composition. These structural properties can be irreversibly controlled by either \emph{in-situ} \cite{Garcia1998, Wasilewski1999} or post-growth \cite{Langbein2004, Ellis2007, Tartakovskii2004} annealing, although with less than deterministic outcomes. Alternatively, reversible \emph{in-situ} manipulation of single particles can be achieved with an electric \cite{Fry2000, Warburton2002}, magnetic \cite{Bayer1999}, or strain \cite{Seidl2006, Plumhof2011, Jons2012} field. Both electric and strain fields have been successfully applied to eliminate the FSS for entangled photon pair generation, \cite{Bennett2010, Gerardot2007, Plumhof2011} to tune QD transitions and cavity modes into resonance \cite{Bonato2011, Midolo2011, Zander2009, Laught2009}, and to obtain resonance between two QDs located on separate chips \cite{Patel2010, Flagg2010}. However, while the behaviour of single particles in QDs in external electric and magnetic fields is well understood, the individual response of electrons and holes in a dot under elastic deformation has yet to be fully characterized. 

Here we investigate the effect of uniaxial stress ($S$) on individual self-assembled InGaAs QDs embedded in a charge tunable device \cite{Warburton2000}. Well-defined Coulomb blockade in the device allows us to apply a perturbative Coulomb-blockade model \cite{Warburton1998, Dalgarno2008} to isolate the effect of strain on the individual electron and hole states and Coulomb interaction energies in single dots. Additionally, we measure the exciton energies, the FSS, and the permanent dipole moment as a function of $dS$. We find that the confinement energies of the electrons ($E_{C}$) and holes ($E_{V}$) can be tuned over a remarkably large range, up to $\approx$ 13~meV in our experimental setup. However, under tensile stress $dE_{C}/dS$ is positive while $dE_{V}/dS$ is negative, leading to an overall modest effect for $dS$ on the exciton recombination energies. Furthermore, the behaviour of different dots under the same uniaxial strain within a small spatial region (< 1~{\textmu}m$^{2}$) varies widely, similar to recent observations by J\"{o}ns \emph{et al.} \cite{Jons2012}. We take advantage of the varying response of individual dots to strain to reversibly tune two QDs on the same chip into resonance, leading to a potentially scalable approach for generating on-chip entanglement between spins in separate QDs.

\begin{figure} [!b]
\includegraphics[scale=.6]{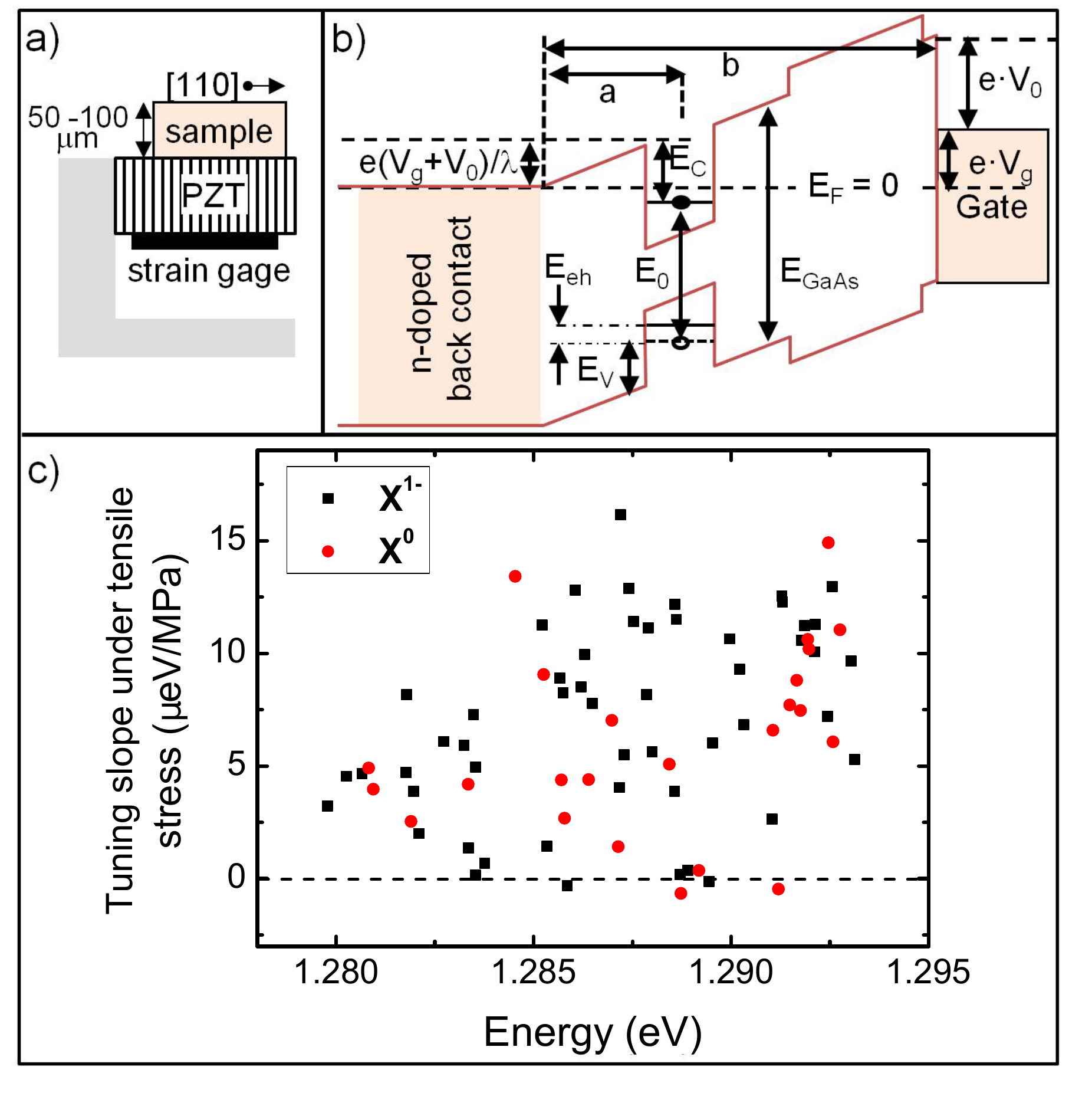}
\caption{Schematic sketches of (a) the experimental setup to apply uniaxial strain and (b) the sample heterostructure. (a) A mechanically thinned ($\approx$ 60~{\textmu}m) GaAs sample is directly glued to a PZT stack with the GaAs [110] direction parallel to the expansion. A strain gage is glued to the bottom of the PZT for monitoring. (b) A sketch of the band diagram for the field effect device identifying the parameters relevant to the Coulomb-blockade model. See \textit{Methods} for the full sample description and Coulomb-blockade model. (c) $dE^{X^{0}}/dS$ and $dE^{X^{1-}}/dS$ as a function of transition energy for several dots in the sample studied. A positive tuning slope refers to a blue-shift under tensile stress.}
\end{figure}

To elastically deform the dots, we directly glue the sample with the [110] crystallographic axis parallel to a piezoelectric lead zirconia titanate (PZT) ceramic stack which allows a bias ($V_{PZT}$) range of $-300 \leq V_{PZT} \leq 300$~V. This corresponds to a calibrated [110] GaAs stress range $-30 \pm 4 \leq S \leq 30 \pm 4$~MPa. The InGaAs QDs are embedded in a charge tunable device allowing deterministic control of the electron occupancy via a DC bias ($V_{Gate}$) (see \textit{Methods}).  In this letter we focus on two exciton states: the neutral (X$^{0}$) and negatively charged (X$^{1-}$) excitons. The results of sweeping $V_{PZT}$ with a constant $V_{Gate}$ clearly show that both X$^{0}$ and X$^{1-}$ transition energies ($E^{X^{0}}$ and $E^{X^{1-}}$, respectively) shift linearly with $V_{PZT}$ but different dots exhibit widely variable tuning slopes. Statistics on the tuning slopes $dE^{X^{0}}/dS$ and $dE^{X^{1-}}/dS$ as a function of nominal energy ($dS = 0$) are shown in Fig.~1c for 78 total transitions. Differences in both $dE^{X^{0}}/dS$ and $dE^{X^{1-}}/dS$ are observed within the same focus at all spatial positions examined, confirming the variable tuning behaviour is due to inherent differences in the QDs. Counter-intuitively, we do not observe a relationship between the transition energies and the tuning slopes. Such varied tuning behaviour is in qualitative agreement with other recent experimental results and theoretical predictions that a dot's response to uniaxial strain is highly dependent on its size, shape, and composition, which varies even for dots with identical transition energies \cite{Jons2012,Singh2010}.  The sample investigated here was grown using the same growth procedure (see \textit{Methods}) as the sample investigated by Seidl \textit{et al.} \cite{Seidl2006} which reported a dot with a significant red-shift under tensile strain. This highlights the extreme sensitivity of the tuning slope (both sign and magnitude) on the dot's shape, size, and composition.

To gain further insight into the response of the dot under uniaxial stress, we uncover the individual behaviour of the electrons and holes in individual dots by investigating Coulomb blockade as a function of $V_{PZT}$. Fig.~2 displays the photoluminescence (PL) (see \textit{Methods}) as a function of $V_{Gate}$ for the X$^{0}$ and X$^{1-}$ transitions of two QDs (labelled QD A and QD B) with similar transition energies at 3 different $V_{PZT}$ values. Under tensile stress, we observe that all four transitions have positive but varied energy tuning slopes (see Table 1). Furthermore, for each dot, with increasing tensile strain the plateau positions shift to more negative $V_{Gate}$ values, the permanent dipole moment increase, the FSS decreases, and the energy difference between X$^{0}$ and X$^{1-}$ increases ($\varDelta E^{X^{0}\rightarrow X^{1-}}$ at $V_{Gate} = V_2$, the most positive $V_{Gate}$ side of the X$^{0}$ plateau). These results are shown in Fig.~3. We can use the PL energies and charging voltages to quantify with sub-meV accuracy \cite{Seidl2005} the relative effect of $S$ on $E_{C}$ and the Coulomb interaction energies $E_{eh}^{11}$ and $E_{ee}^{21}$ (using the notation $E_{ab}^{\alpha \beta}$ where $ab$ identifies the type of Coulomb interaction, $ee$ for electron-electron and $eh$ for electron-hole, $\alpha$ refers to the number of electrons, and $\beta$ the number of holes in the dot). We use the highly accurate values $dE_{C}/dS$, $dE^{X^{0}}/dS$, and $dE_{eh}^{11}/dS$ to estimate $dE_{V}/dS$ (see \textit{Methods}). The results shown in Fig.~3 and Table 1 establish a complete picture of the effect of $dS$ on the single particles and exciton complexes. 

\begin{figure} [!t]
\includegraphics[scale=.8]{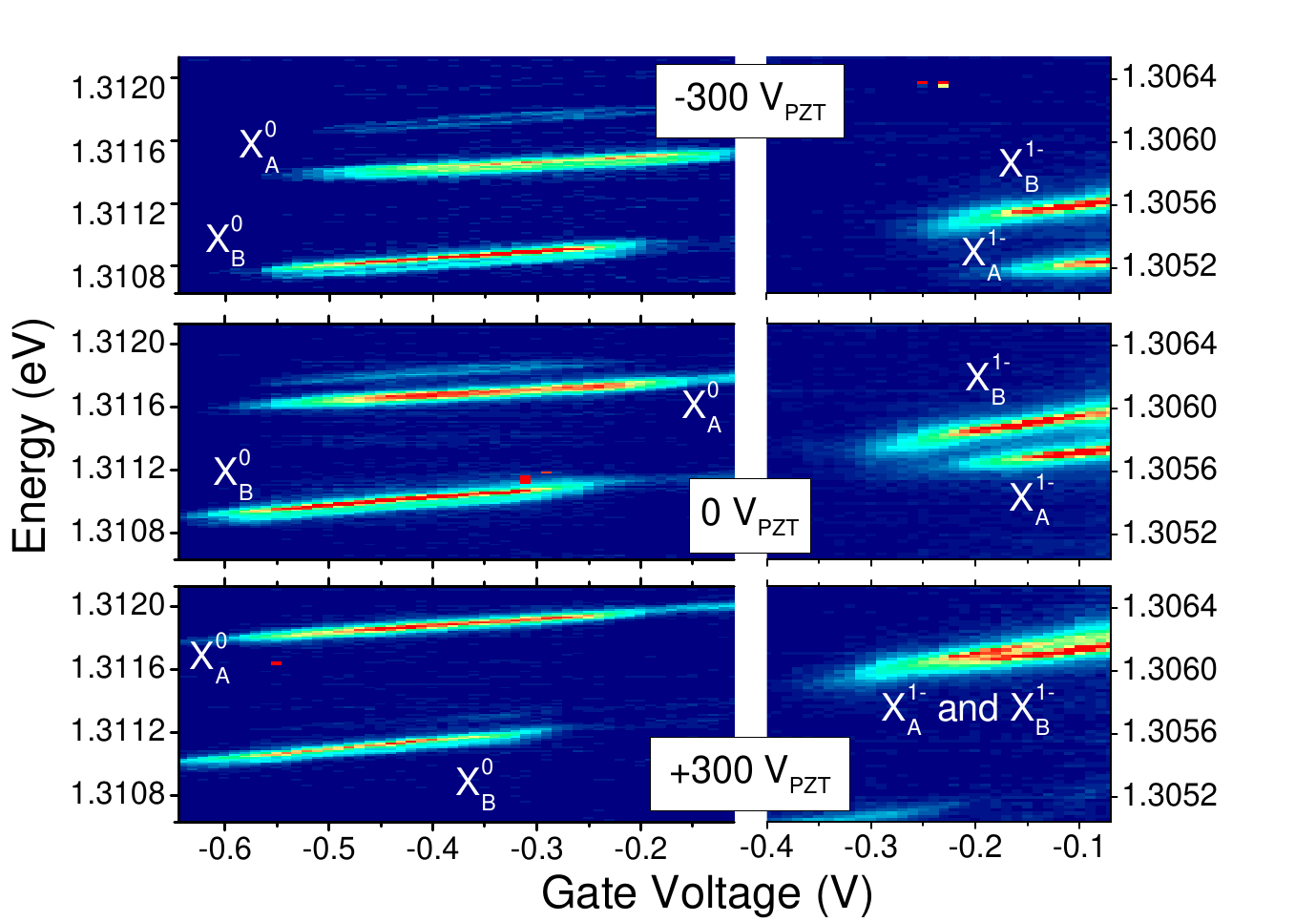}
\caption{PL from QDs A and B as a function of sample gate bias for three different amounts of applied strain. As $V_{PZT}$ = -300~V $\rightarrow$ $V_{PZT}$ = +300~V each transition blue-shifts, $V_{Gate}$ for the charge plateau decreases, and the Stark shift increases. Due to the different response of the two dots to strain, $X_{A}^{1-}$ and $X_{B}^{1-}$ merge together at $V_{PZT}$ = +300~V.}
\end{figure}

\begin{table} [!b]
  \caption{Tuning slopes (in units {\textmu}eV/MPa) for the exciton, single-particle, and fine-structure splitting energies for four different quantum dots under tensile stress.}
  \label{tbl:one}
  \begin{tabular}{ccccccc}
    \hline
    QD & $dE^{X^{0}}/dS$ & $dE^{X^{1-}}/dS$  & $d\varDelta E^{X^{0} \rightarrow X^{1-}}/dS$  & $dE_{C}/dS$  & $dE_{V}/dS$ & $dFSS/dS$ \\
    \hline
    A  & 7.39 & 7.86 & -0.48 & 203 & -211 & -0.41\\
    B  & 7.20 & 8.63 & -1.44 & 226 & -233 & -0.36\\
    C  & 4.19 & 4.50 & -0.32 & 230 & -234 & -0.28\\
    D  & 7.01 & 7.89 & -0.87 & 186 & -193 & -0.47\\
    \hline
  \end{tabular}
\end{table}

A key result revealed by the values for $dE_{C}/dS$ and $dE_{V}/dS$ is that the single electron and hole states both red-shift significantly ($\approx 0.2$~meV/MPa), but $d\varDelta E^{X^{0} \rightarrow X^{1-}}/dS$ is small ($\approx -1$~{\textmu}eV/MPa), indicating that the Coulomb interaction energies are modestly affected by $dS$. We note that values determined by the Coulomb-blockade model for $dE_{eh}^{11}/dS$ and $dE_{ee}^{21}/dS$ are less than 20~{\textmu}eV/MPa, the model's accuracy limit. Hence, $dE^{X^{0}}/dS$ and $dE^{X^{1-}}/dS$ are determined predominantly by the considerable change in the single particle energies. The many-body effects are observed to be much weaker. Theoretical modelling \cite{Singh2010, Jons2012} of dots with varying size, shape, and composition agree with the range of tuning slopes for exciton energies we observe. Combining the experimental and theoretical results yields a crucial insight: the precise balance of $dE_{C}/dS$ and $dE_{V}/dS$ determines the tuning of the exciton energies and is highly sensitive to the dot's inherent structural properties, not solely on the initial confinement potentials.

\begin{figure} [!b]
\includegraphics[scale=1]{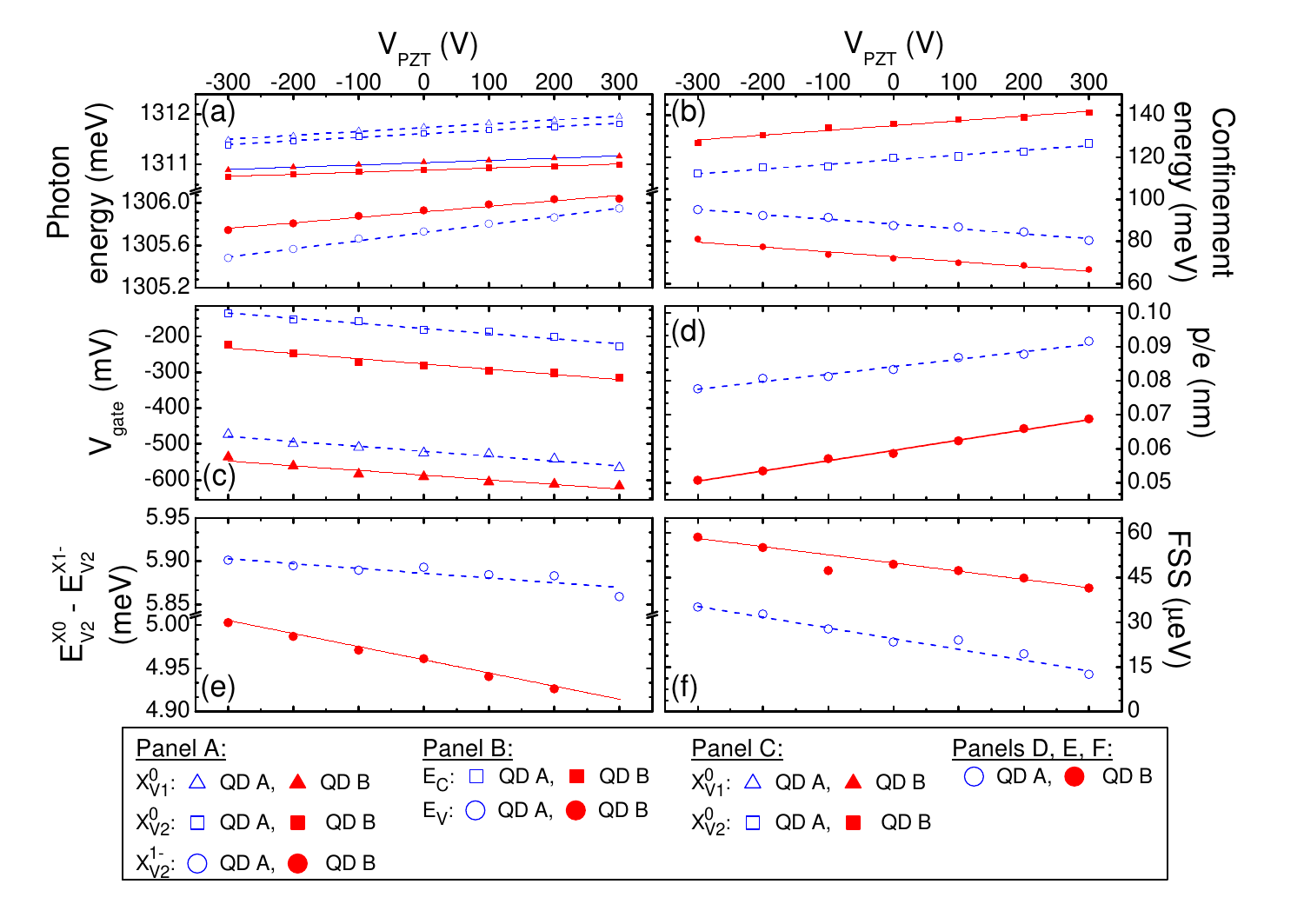}
\caption{The response of QDs A and B as a function of uniaxial strain for several properties. $V_{1}$ refers to the most negative $V_{Gate}$ edge of the $X^{0}$ plateaus and $V_{2}$ to the most positive (negative) $V_{Gate}$ edge of the $X^{0}$ ($X^{1-}$) plateaus in Fig.~2.}
\end{figure}

The change of the single exciton transition energy in response to a vertical electric field, i.e. the quantum confined Stark shift, also varies linearly with $dS$. The Stark effect is given by $E(F) = E_{0} - p \cdot F + \beta \cdot F^{2}$, where $F$ is the electric field, $E_{0}$ is the zero-field transition energy, $p$ is the intrinsic vertical dipole moment which reflects the zero-field separation ($p/e$) of the electron and hole wave functions, and $\beta$ is the polarizability. For $X^{0}$ and $X^{1-}$ transitions in this field-effect structure, the quadratic term of the Stark effect is negligible and linear plateaus are generally observed. Fig.~3d shows that $p$ increases with increasing tensile stress, even though the change in confinement energies mostly cancel each other.  Notably, the values for $p/e$ at $V_{PZT}$ = 0~V are less than previously reported, \cite{Fry2000, Warburton2002} possibly due to pre-strain on the sample arising during sample cool-down.

For each dot investigated the FSS also changes in response to $dS$ (Fig.~3f and Table~1). $dFSS/dS$ quantitatively agrees with theoretical predictions. \cite{Gong2011} The maximum FSS tuning range observed for dots investigated in our experimental setup is $\approx$ 28~{\textmu}eV and we can observe dots for which the FSS vanishes within the resolution limit (e.g. QD1 at $V_{PZT}$ = +300~V in Fig.~4).  All FSS tuning dependencies remain linear with $dS$, and we observed a strain-induced rotation in the polarization eigenstates of some dots of about 10 degrees. Compared to electric field tuning techniques, a huge benefit of strain tuning is a constant oscillator strength. We confirm the PL intensity is constant as $V_{PZT}$ is swept provided that the QD spatial position is re-adjusted. 

\begin{figure} [!t]
\includegraphics[scale=0.4]{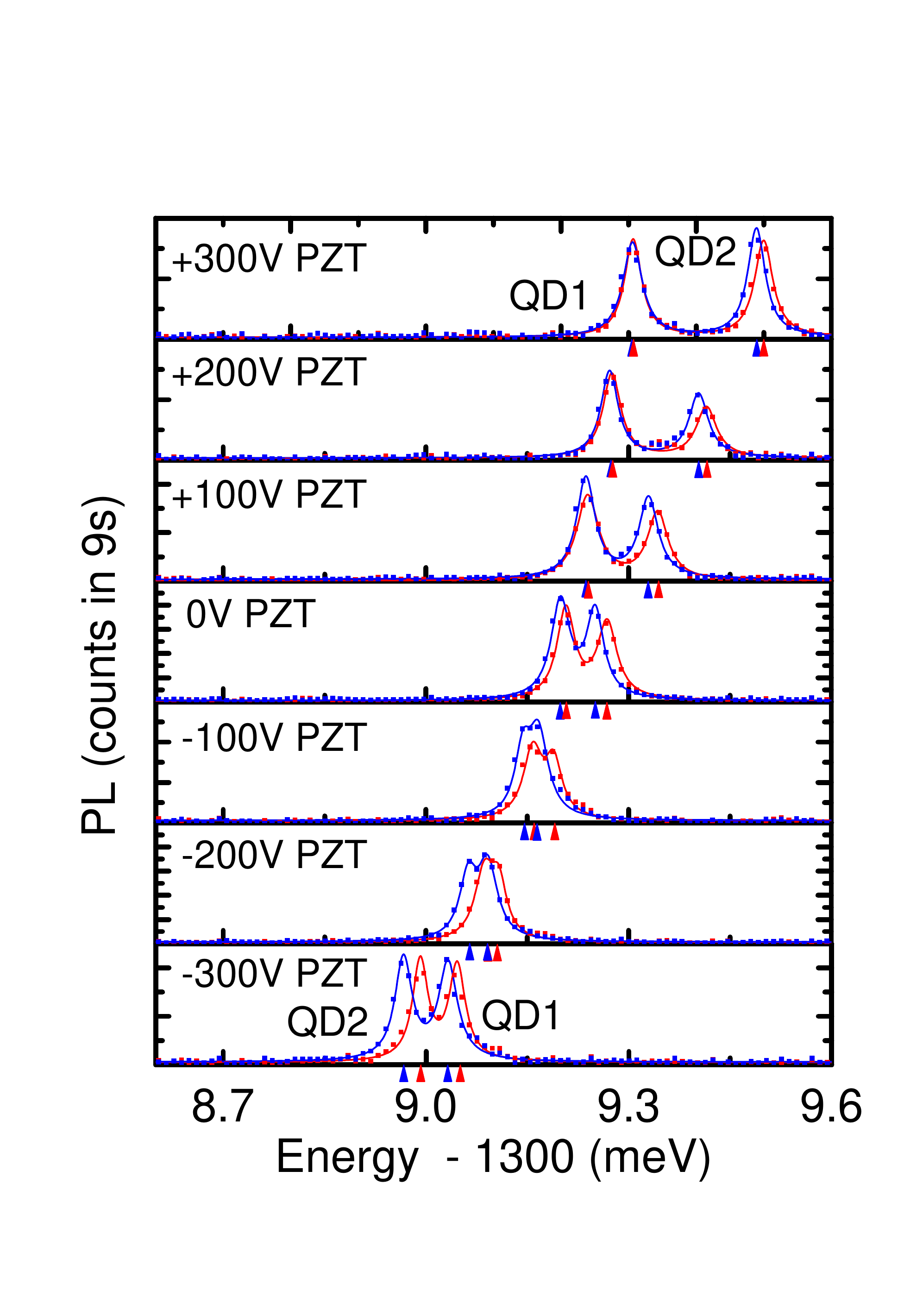}
\caption{Strain tuning two QDs into resonance within the same focal position. QD2 exhibits a larger response to the strain than QD1 allowing tuning through degeneracy for each transition between the two QDs. $E_{x}$ (blue) and $E_{y}$ (red) show the data points and double-peak Lorentzian fits (solid lines) for the orthogonal polarizations in the basis of the FSS for both dots. The triangles beneath the spectra identify the center peak positions from the fits. In addition to the center peak positions of QD1 and QD2, the FSS for each dot is affected by the uniaxial strain. Notably, for QD1 the FSS < 10~{\textmu}eV at $V_{PZT}$ = +300~V and cannot be distinguished. While the spatial position is re-adjusted for each $V_{PZT}$, the counts for each dot still fluctuate due to the strain. [scale: minor tick marks = 200~cts]}
\label{Fig5}
\end{figure}

Finally, we demonstrate it is possible to exploit the variable tuning slopes to reversibly and deterministically tune transitions in separate dots through degeneracy. One such example is shown in Fig.~4 for the $X^{0}$ transitions from two dots (labelled QD1 and QD2). At $V_{PZT} = +300$~V, the energy difference ($\varDelta E^{X^{0}}_{1-2}$) between the two dots is -190~{\textmu}eV (where $E^{X^{0}}_{1}$ and $E^{X^{0}}_{2}$ are defined as the middle of the FSS). However, QD1 has a smaller tuning slope than QD2, so that $\Delta E^{X^{0}}_{1-2}$ = +67~{\textmu}eV at $V_{PZT} = -300$~V. Each transition from each dot becomes degenerate at a separate $V_{PZT}$ value. For instance, at $V_{PZT}$ $\approx$ -200~V, the $y$-polarized transition from QD1 crosses the $x$-polarized transition from QD2. This highlights the potential for a scalable approach to achieve indistinguishable photons from multiple independent QDs on the same chip and in the same cryostat. 

In summary, we utilize electro-elastic functionality to reversibly tune single particles confined within self-assembled QDs. Insight into the response of the single electron and hole states under uniaxial strain is uncovered using a perturbative Coulomb-blockade model. We find that while the electron (hole) confinement energies are substantially blue (red)-shifted under tensile stress, tuning of the many-body interaction energies and FSS is more modest. Due to the opposite response of the single electron and hole states, the resulting exciton behaviour varies substantially within the dot population. The relative magnitudes of the strain tuning of single electrons and holes are determined by each individual dot's inherent size, shape, and composition. When combined with the excitonic spectra \cite{Ediger2007}, electro-elastic spectroscopy offers a powerful non-destructive technique to determine the structure and symmetry of single QDs, an emerging concept in probing and designing quantum structures. \cite{Mlinar2009, Chekhovich2011}

\section{Methods}%
\subsection{Experimental setup}%
To elastically deform the dots, we follow the procedure of Shayegun \textit{et al.} \cite{shayegan2003} and directly glue a mechanically polished GaAs sample ($\approx$ 50 - 100~{\textmu}m thickness) to a PZT ceramic stack (Piezomechanic 150/10x10/7) using a two component epoxy (UHU Plus Endfest 300) as depicted in Fig.~1a. The stretching axis of the PZT is parallel to the [110] GaAs crystallographic direction. A strain gage is glued to the bottom of the PZT stack for monitoring. At the experiment temperature (T = 4.2~K), a maximum bias $V_{PZT}$ = $\pm$ 300~V can be applied. In the setup, a positive (negative) $V_{PZT}$ corresponds to expansion (contraction) of the PZT stack and tension (compression) of the GaAs. Creep and hysteresis is minimal in this setup ($\delta L/L \propto V_{PZT}$). Due to strain, as $V_{PZT}$ is swept the single QDs move in and out of the excitation / collection focus, which has a resolution of 755 $\pm$ 10 nm. \cite{Gerardot2007b} $\delta L$ at the QD position is calibrated from the observed FWHM of the dots' intensities, $V_{PZT}^{FWHM}$ = 250 $\pm$ 16~V.  This leads to a maximum strain change over $V_{PZT}$ = 600~V of $\delta L/L$ = (4.9 $\pm$ 0.3)$\times 10^{-4}$.  Using $Y = 121.3$ GPA as the Young's modulus of GaAs along the [110] direction, we obtain $\left| S_{maximum} \right|$ = 30 $\pm$ 4~MPa. $S$ calculated from the changing spatial position of the dots is in rough agreement with the strain gage measurement. 

We perform PL using an 826~nm excitation laser. The full width at half maximum (FWHM) of individual dot peaks is measured to be 30~{\textmu}eV with our spectrometer, but the peak positions can be determined with < 10~{\textmu}eV accuracy using a Lorentzian fitting procedure. The FSS is determined via polarization dependent PL using a fixed linear polarizer and a rotating half waveplate. 

\subsection{Sample details}%
The InGaAs QDs are grown on a (100) GaAs wafer by the partially capped island technique \cite{Garcia1998}, in which self-assembled InAs islands are partially capped with 30~\AA{} GaAs and annealed for 30~s at the QD growth temperature before completely covering the structure with GaAs. This growth procedure introduces significant shape, strain, and composition changes in the QDs, effectively shifting the ground state transition energies to $\approx$ 950 $\pm$ 25~nm. The highly alloyed dots are embedded in a charge tunable device (Fig.~1b). A 25~nm GaAs tunnel barrier separates the dots from a grounded 20~nm thick n+ GaAs layer (doping $\approx 10^{18}$ cm$^{-3}$). To isolate the QD from electric fields from the PZT, the ground of the PZT stack and the back gate of the sample are connected. A 30~nm capping layer separates the dots from a 100~nm thick AlAs/GaAs blocking barrier. A voltage ($V_{Gate}$) applied to a semitransparent NiCr Schottky gate on the top surface shifts the QD energy levels with respect to the Fermi level in the grounded back gate, allowing deterministic electron charging and pronounced Coulomb blockade.

\subsection{Coulomb-blockade model}%
We assume the dot is in the strong confinement limit, e.g. the electron and hole wave functions are determined by the confining potential and only slightly perturbed by the Coulomb interactions, and use an established perturbative Coulomb-blockade model \cite{Warburton1998,Dalgarno2008,Seidl2005}. As shown in Fig.~1b and Fig.~3, the input parameters for the model include: the tunnel barrier thickness $a$; the back gate to surface Schottky gate distance, $b$; the built-in voltage of the Schottky contact, $V_{0}$ ($V_{0} = 0.62$ V); the electrostatic energy due to an image charge in the back contact, $E_{i}$ ($E_{i} = -1.1$ meV) for one electron; the separation between $E_{C}$ and $E_{V}$, $E_{0}$; the $V_{Gate}$ at which a single electron tunnels into the dot (the left edge of the X$^{0}$ plateau), $V_{1}$; the $V_{Gate}$ at which a second electron tunnels into the dot (the right edge of the X$^{0}$ plateau), $V_{2}$; the X$^{0}$ photon energy at $V_{1}$, $E^{X^{0}}$; and the difference in X$^{0}$ and X$^{1-}$ photon energies at $V_{2}$, $\varDelta E^{X^{0} \rightarrow X^{1-}}$. The Fermi energy is set to 0 ($E_{F} \equiv 0$) so the potential at the dot is $e(V_{0}-V_{Gate})/\lambda$, where $\lambda = b/a = 6.45$. Including Coulomb interactions, the relevant photon energies are $E^{X^{0}} = E_{0} - E_{eh}^{11}$ and $E^{X^{1-}} = E_{0} - 2E_{eh}^{11} + E_{ee}^{21}$ and the gate voltage extent of the $X^{0}$ plateau is $e(V_{2} - V_{1})/\lambda = E_{ee}^{21} - 2E_{i}$. To obtain $E_{C}$, we first find $E_{ee}^{21}$ from the experimentally determined values $V_{1}$ and $V_{2}$. $E_{eh}^{11}$ can then be experimentally found from $\varDelta E^{X^{0} \rightarrow X^{1-}}$, which allows us to determine $E_{C}$ based on the gate voltage at which a single electron tunnels into the dot ($V_{1}$). Unfortunately, the charging voltages $V_{1}$ and $V_{2}$ measured via PL are shifted together as $V_{Gate}$ is intensity dependent down to extremely low excitation powers \cite{Seidl2005} because of a charge storage effect in the device. However, under constant excitation powers we can determine the \textit{relative} changes in the charging voltages due to $dS$ with very high accuracy. We note that the value $dE_{C}/dS$ is the conduction band offset (CBO) of the single electron state in the dot including the change in the barrier offset due to $dS$.  Finally, we can obtain the confinement energy for holes in the valence band: $E_{V} = E_{GaAs} - E_{0} - E_{C}$, where $E_{GaAs}$ is the GaAs band-gap, which we set to 1.5187~eV. We set $E_{GaAs}$ to this constant value as the change in the valence band offset (VBO) of the barrier due to $dS$ and the relative contributions of the conduction-band and valence band to $dE_{GaAs}/dS$ are not established. Hence, the values we report for $dE_{C}/dS$ are robust measurements including the change in $dCBO/dS$, whereas the value for $dE_{V}/dS$ is an estimate which does not include $dVBO/dS$. Nevertheless, by combining the robust values of $dE_{C}/dS$, $dE^{X^{0}}/dS$, and $dE^{X^{1-}}/dS$, a complete understanding of the quantum dot's response to $dS$ is obtained.

\acknowledgement
We thank K. Karrai and R. J. Warburton for helpful discussions, and acknowledge funding from EPSRC and the Royal Society.

\end{document}